\documentclass
[superscriptaddress,secnumarabic,amssymb,amsmath,nobibnotes,aps,prd,showkeys,showpacs,nofootinbib,onecolumn]{revtex4}%
\usepackage{graphicx}
\usepackage{epsf}
\usepackage{bm}
\usepackage{amsmath}
\usepackage{amsfonts}
\usepackage{amssymb}%
\providecommand{\U}[1]{\protect\rule{.1in}{.1in}}

\newcommand{\be}{\begin{equation}}
\newcommand{\ee}{\end{equation}}

\newcommand{\mincir}{\raise
-3.truept\hbox{\rlap{\hbox{$\sim$}}\raise4.truept\hbox{$<$}\ }}
\newcommand{\magcir}{\raise
-3.truept\hbox{\rlap{\hbox{$\sim$}}\raise4.truept\hbox{$>$}\ }}

\begin{document}
\title{Similarity Inner Solutions for the Pulsar Equation}
\author{Andronikos Paliathanasis}
\email{anpaliat@phys.uoa.gr}
\affiliation{Institute of Systems Science, Durban University of Technology, PO Box 1334,
Durban 4000, Republic of South Africa}
\begin{abstract}
Lie symmetries are applied to classify the source of the magnetic field for
the Pulsar equation near to the surface of the neutron star. We find that
there are six possible different admitted Lie algebras. We apply the
corresponding Lie invariants to reduce the Pulsar equation close to the
surface to an ordinary differential equation. This equation is solved
either with the use of Lie symmetries or the application of the ARS algorithm for
singularity analysis to write the analytic solution as a Laurent expansion.
These solutions are called inner solutions.


\end{abstract}
\maketitle
\date{\today}

\section{Introduction}

\label{introduction}

Pulsars are one of the most impressive observable celestial objects in the
sky. They are assumed to be rotating neutron stars which emit radio signals.
However, their importance follows from the fact that they are physical laboratories which
provide extreme conditions of strong magnetic fields which cannot be
reproduced on Earth.

The structure of the strong magnetic fields in a Pulsar is described by a
scalar function which satisfies the elliptic second-order partial
differential equation \cite{pulsar1,pulsar2},%
\begin{equation}
\left(  1-x^{2}\right)  \left(  \Psi_{,xx}+\Psi_{,zz}\right)  -\frac{1+x^{2}%
}{x}\Psi_{,x}+F\left(  \Psi\right)  =0, \label{eq.01}%
\end{equation}
where the singularities at $x=0$ and $x=1$ represent the centre of the star,
($x$ is the radius coordinate) and the surface of the pulsar is
located at $x=1$. Function $F\left(  \Psi\right)  $ is related to the profile of the
magnetic field for the polar coordinate \cite{pulsar1}. Equation (\ref{eq.01})
is also known as the relativistic force-free Grad-Shafranov equation
\cite{ppl1}.

In order to arrive at such a simple scalar equation, (\ref{eq.01}), for the
magnetic field, various Ans\"atze have been assumed for the physical state of
the star. In particular it has been assumed that \cite{pulsar1}: (a) the
system is axisymmetric and time-independent; (b) the electrons and the ions
have a well-defined velocity and density; (c) there are no gravitational or
particle collision effects; (d) inertial forces have been considered and (e)
it is assumed that the surface of the uniformly rotating star is a perfect conductor.

Because of the nonlinearity and the existence of the two singular points, the
Pulsar equation, (\ref{eq.01}), cannot be integrated in general and only few
solutions are known in the literature.\ Originally, an asymptotic analytical
solution which describes the magnetic field near to the surface of the star was
presented by Michel in \cite{pulsar2}. This was also the main inspiration for
the recent works of Uzdensky \cite{pl3} and Gruzinov \cite{pl4}. In
\cite{pl3} an interesting discussion of the physical state of the boundary
conditions is given. However, numerical solutions which describe the global
evolution of the Pulsar equation have been presented in the literature. One of
the first numerical force-free solutions was derived by Contopoulos et al.
\cite{pl5}, while other numerical solutions can be found in
\cite{pl6,pl7,pl8,pl9} and references therein.

In this work we are interested to apply the powerful method of Lie
symmetries~\cite{Stephani,Bluman} in order to study the existence of invariant
solutions for the Pulsar equation near to the singularity, $x=1,~$and to find
analytical asymptotic solutions, the so-called similarity solutions. In
particular, we classify the source of the magnetic field, i.e. function
$F\left(  \Psi\right)  $, such that the Pulsar equation, near to the
singularity, $x=1$, be invariant under the action of one-parameter point
transformations. This kind of classification was firstly introduced by Ovsiannikov
\cite{ovsiannikov} and has been applied to various physical systems for the
determination of new analytical solutions, for instance see
\cite{ref1,ref2,ref3,ref4,ref5,ref6,ref7,ref8,ref9,ref10,qm1,qm2,qm3,qm4} and
references therein, for various applications of the Lie symmetry
classification in Physics.

The novelty of Lie symmetries is that symmetries can be used to define
invariant surfaces and to reduce the number of dependent variables -- for
partial differential equations -- or to reduce the order of the differential
equation for ordinary differential equations. Hence new integrable systems
can be constructed and new analytical solution to be determined. The plan of
the paper follows.

In Section \ref{sec2a} the basic properties and definitions for the Lie
(point) symmetries of differential equations are presented. In the same
Section we perform the classification of the Lie symmetries of the pulsar
equation near to the surface of the star and we find that there are six
different admitted groups of point-transformations which leave the pulsar
equation invariant for six different functional form of the source, $F\left(
\Psi\right)  $. The context of singularity analysis is discussed which is used
in subsequent Sections to prove the integrability of some of the reduced
differential equations. The application of the Lie symmetries and the
determination of the similarity inner solutions is performed in Section
\ref{sec3a}. New asymptotic analytic solutions near to the surface of the star
are presented. Finally in Section \ref{sec4a} we discuss our results and we
draw our conclusions.

\section{Lie symmetry analysis}

\label{sec2a}

For the convenience of the reader we present the
basic properties and definitions of Lie point symmetries of differential
equations and more specifically we discuss the case of second-order
differential equations of the form $\mathbf{A\equiv}A\left(  x^{k}%
,\Psi,\Psi_{,i},\Psi_{,ij}\right)  =0$, where $x^{k}$ are the independent
variables and $\Psi$ is the dependent variable with first derivative
$\frac{\partial\Psi}{\partial x^{i}}=\Psi_{,i}$.

Let
\begin{equation}
X=\xi^{i}\left(  x^{k},u\right)  \partial_{i}+\eta\left(  x^{k},\Psi\right)
\partial_{u},\label{go.10}%
\end{equation}
be the generator of the local infinitesimal one-parameter point transformation,
\
\begin{align}
\bar{x}^{k} &  =x^{k}+\varepsilon\xi^{i}\left(  x^{k},\Psi\right)  ,\\
\bar{\eta} &  =\eta+\varepsilon\eta\left(  x^{k},\Psi\right)  .
\end{align}

Then $X$ is called a Lie symmetry for the differential equation, $\mathbf{A}$,
iff
\begin{equation}
X^{\left[  2\right]  }A=\lambda A\label{go.11}%
\end{equation}
in which $X^{\left[  2\right]  }$ is called the second prolongation/extension
in the jet-space and is defined as%
\begin{equation}
X^{\left[  1\right]  }=X+\left(  D_{i}\eta-\Psi_{,k}D_{i}\xi^{k}\right)
\partial_{\Psi_{,i}}+\left(  D_{i}\eta_{j}^{\left[  i\right]  }-\Psi_{jk}%
D_{i}\xi^{k}\right)  \partial_{\Psi_{,ij}}.\label{go.13}%
\end{equation}

The novelty of Lie symmetries is that they can be used to determine similarity
transformations, i.e. differential transformations where the number of
independent variables is reduced \cite{Bluman}. The similarity transformation
is calculated with the use of the associated Lagrange's system,
\begin{equation}
\frac{dx^{i}}{\xi^{i}}=\frac{du}{\Psi}=\frac{d\Psi_{i}}{\Psi_{\left[
i\right]  }}=...=\frac{d\Psi_{ij..i_{n}}}{\Psi_{\left[  ij...i_{n}\right]  }}.%
\end{equation}

Solutions of partial differential equations which are derived with the
application of Lie invariants are called similarity solutions. In this
specific work we use the Lie symmetries to reduce the Pulsar equation to a second-order
differential equation. For this equation we shall analytic
solutions by using the symmetry approach and, if we fail, we apply the
singularity analysis.

\subsection{Singularity analysis}

Singularity analysis is another powerful mathematical method which is
applied to study the integrability of differential equations and to present the
solutions of differential equations in algebraic form, in particular by using
Laurent expansions around a movable singularity.

Singularity analysis is also known as the Painlev\'{e} Test
\cite{Painleve1,Painleve2,Painleve3,Painleve4} and has been applied in various
problems for the study of integrability of \ given differential equations.

Ablowiz, Ramani and Segur \cite{Abl1,Abl2,Abl3} systematized the Painlev\'{e}
Test in a simple algorithm, also known as the ARS algorithm. The main feature of
the ARS\ algorithm is its simplicity. It consists of three main algebraic
stems: (a) determination of leading-order behaviour; (b) determination of
resonances and (c) consistency of Laurent expansion. For every step of the
algorithm there are various criteria which should be applied, these criteria
are summarized in the review of Ramani et al. \cite{buntis}.

If a given differential equation passes the three steps of the ARS algorithm,
then we conclude that the given differential equation is algebraically integrable.
However, should the differential equation fail the ARS
algorithm, we cannot make a conclusion about the integrability of the differential
equation. While the ARS algorithm is straightforward on its application, one
of the main disadvantages is that it depends upon the coordinates in which the
given equation is defined, for a recent discussion we refer the reader to
\cite{anleach}.

\subsection{Pulsar equation near to the singularity}

We define the new coordinate, $y=x-1$, in order to move the surface of the
star to $y=0$. It follow, $y>0$, when $x>1$ and $y<0$ , when $x<1$. In the new
coordinates the Pulsar equation (\ref{eq.01}) becomes%
\begin{equation}
y\left(  2+y\right)  \left(  \Psi_{,yy}+\Psi_{,zz}\right)  +\frac{\left(
2+2y+y^{2}\right)  }{1+y}\Psi_{,y}-F\left(  \Psi\right)  =0. \label{eq.02}%
\end{equation}

Near to the surface with $y\simeq0$, (\ref{eq.02}) is
approximated by the simpler form \cite{pulsar2}
\begin{equation}
2y\left(  \Psi_{,yy}+\Psi_{,zz}\right)  +2\Psi_{,y}-F\left(  \Psi\right)  =0.
\label{eq.02a}%
\end{equation}
This is the equation which Michel \cite{pulsar2} used to find the first analytical
expression for the force-free magnetosphere and inspired the later works of
\cite{pl3,pl4}. Equation (\ref{eq.02a}) is the one that we use to perform the
symmetry classification.

Moreover, we follow \cite{pl3,pl4} and we work on the polar-like coordinates
\begin{equation}
y=r\sin\theta~,~z=r\cos\theta, \label{p.03}%
\end{equation}
where equation (\ref{eq.02a}) takes the form%
\begin{equation}
2r\sin\theta\left(  \Psi_{,rr}+\frac{1}{r^{2}}\Psi_{,\theta\theta}\right)
+2\left(  2\sin\theta+\frac{\cos\theta}{r}\right)  \Psi_{,r}-F\left(
\Psi\right)  =0. \label{pl.04}%
\end{equation}
Hence the surface is indicated when $r=0$ or $\theta=0$. We continue with the
classification of the sources, $F\left(  \Psi\right)  $, such that equation
(\ref{pl.04}) be invariant under one-parameter point transformations, i.e.
Lie symmetries exist, while in the following section we discuss the
application of the Lie symmetries by performing reduction of the equation with
the use of the Lie invariants.

\subsection{Symmetry classification}

For the second-order differential equation (\ref{pl.04}) the symmetry
condition (\ref{go.11}) provides that for arbitrary function, $F_{A}\left(
\Psi\right)  =F\left(  \Psi\right)  $, the differential equation admit the
unique symmetry vector
\[
Y=\cos\theta\partial_{r}-\frac{\sin\theta}{r}\partial_{\theta}.
\]
That vector field corresponds to the translation symmetry, $\partial_{z}$, in
the original coordinates, for equation (\ref{eq.02}) which is also a symmetry
of equation (\ref{eq.01}). Reduction with the use of the symmetry vector
$Y$ leads to solutions which are independent of the $z-$direction and are
not of special interest.

However, for specific functions, $F\left(  \Psi\right)  $, the differential
equation (\ref{pl.04}) can be invariant under a higher dimensional
Lie algebra. In particular we find five different cases:

\begin{itemize}
\item When the source is constant, i.e. $F_{B}\left(  \Psi\right)  =F_{0}$,
the differential equation (\ref{pl.04}) admits four plus infinity symmetries,
these are.%
\[
X_{1}=\partial_{r}~,~X_{2}=Y~,~X_{3}=r^{2}\cos\theta\partial_{r}+r\sin
\theta\partial_{x}-r\cos\theta~\Psi\partial_{\Psi}%
\]%
\begin{equation}
X_{4}=\Psi\partial_{\Psi}~,~X_{\infty}=b\left(  r,\theta\right)
\partial_{\Psi},%
\end{equation}
where $b\left(  r,\theta\right)  $ is a solution of equation (\ref{pl.04}).
The last two symmetries, i.e. $X_{4}$ and $X_{\infty},$ denote the linearity
of equation (\ref{pl.04}). The Lie Brackets of the admitted algebra are given
in Table \ref{tac1}
\end{itemize}

~%
\begin{table}[tbp] \centering
\caption{Lie Brackets of the admitted Lie symmetries for the free pulsar
equation \ref{pl.04}}$%
\begin{tabular}
[c]{c|cccc}%
$\left[  \cdot,\cdot\right]  $ & $X_{1}$ & $X_{2}$ & $X_{3}$ & $X_{4}$\\\hline
$X_{1}$ & $0$ & $0$ & $0$ & $0$\\
$X_{2}$ & $0$ & $0$ & $-X_{3}$ & $X_{4}$\\
$X_{3}$ & $0$ & $X_{3}$ & $0$ & $2X_{2}-X_{1}$\\
$X_{4}$ & $0$ & $-X_{4}$ & $-2X_{2}+X_{1}$ & $0$%
\end{tabular}
$\label{tac1}%
\end{table}%

\begin{itemize}
\item For linear source, $F_{C}\left(  \Psi\right)  =F_{1}\Psi,~$the
differential equation admits two plus infinity symmetries, those
are~$X_{2}~,~X_{4}~$and$~X_{\infty}.$

\item Moreover, for the power-law source, $F_{F}\left(  \Psi\right)
=F_{1}\Psi^{\frac{1}{N}+1},~N\neq0,\frac{1}{2},-1,~$Pulsar equation near to
the surface admits two Lie point symmetries, these are%
\begin{equation}
X_{2}~,~X_{\left(  N\right)  }=r\partial_{r}-N\Psi\partial_{\Psi}%
\end{equation}
with Lie Bracket $\left[  X_{2},X_{\left(  N\right)  }\right]  =X_{2}$.

\item In the special case for which$~N=\frac{1}{2}$ in the latter case, or
$F_{E}\left(  \Psi\right)  =F_{1}\Psi^{3},~$the Pulsar equation (\ref{pl.04})
is invariant under a three-dimensional Lie algebra with elements the vector
fields $X_{2},~X_{\left(  1/2\right)  },~X_{3},~$and Lie Brackets as are
presented in Table \ref{tac2}.
\end{itemize}

%

\begin{table}[tbp] \centering
\caption{Lie Brackets of the admitted Lie symmetries for the pulsar equation
with cubic-law force}$%
\begin{tabular}
[c]{c|ccc}%
$\left[  \cdot,\cdot\right]  $ & $X_{2}$ & $X_{\left(  1/2\right)  }$ &
$X_{3}$\\\hline
$X_{2}$ & $0$ & $-X_{(1/2)}$ & $X_{3}$\\
$X_{\left(  N\right)  }$ & $X_{(1/2)}$ & $0$ & $2X_{2}$\\
$X_{3}$ & $-X_{3}$ & $-2X_{2}$ & $0$%
\end{tabular}
$\label{tac2}%
\end{table}%

\begin{itemize}
\item Finally, for the exponential source, $F_{F}\left(  \Psi\right)
=F_{1}e^{-\frac{1}{C}\Psi}$,~$C\neq0,$ the Pulsar equation admits two Lie
point~symmetries,~%
\[
X_{2}~,~\bar{X}_{\left(  C\right)  }=r\partial_{r}+C\partial_{\Psi}\text{.}%
\]
with Lie Bracket $\left[  X_{2},\bar{X}_{\left(  C\right)  }\right]  =X_{2}$.
We mention that the exponential-lie source was introduced in \cite{ppl1} as a jet model.
\end{itemize}

We continue with the application of the Lie symmetry vectors to determine
analytical solutions of the Pulsar equation (\ref{pl.04}). The solutions that
we determine are valid as first approximations of the general solution
near to the surface of the star. In particular, near to the surface of the
star, $y\simeq0$, the differential equation can be seen as a singular
pertubative equation and the theory of singular perturbative differential
equations \cite{sper1,sper2} can be applied in order to justify the
approximation of the analytical solution. The solutions near to the point
$y\simeq0$ are called inner solutions \cite{sper1}.

\section{Similarity solutions}

\label{sec3a}

As we discussed in the previous Section, for every Lie symmetry we can define
a surface where the solution is independent of one of the variables, that is,
define similarity variables.

For arbitrary source, $F\left(  \Psi\right)  $, from the vector field
$\partial_{z}$ the invariant solution is the one where $\Psi\left(
y,z\right)  =\Psi\left(  y\right)  $ and the resulting differential equation
is the ordinary differential equation
\begin{equation}
2y\Psi_{,yy}+2\Psi_{,y}-F\left(  \Psi\right)  =0. \label{sc.01}%
\end{equation}
That is not a solution of special interest.  Hence we proceed with our analysis by
using the remainder of the symmetry vectors.

\subsection{Invariant solutions for constant source}

The case of constant source also covers the free-source problem when
$F_{0}=0$. Indeed in equation (\ref{eq.02a}) for $F\left(  \Psi\right)
=F_{0}$ we can replace $\Psi\rightarrow\Psi+\frac{F_{0}}{2}y$.  Then the
source-free case follows. From table \ref{tac1} it follows that there are
four possible reductions which we can perform. They are: (a) reduction with the
symmetry vector $X_{1}+\mu X_{4}$;~(b) reduction with $X_{2}+\mu X_{4}$;~(c)
reduction with $X_{3}+\mu X_{4}$; and (d) reduction with $X_{3}+\mu X_{2}$.
\ For each of these reductions the reduced equation is a linear second-order
differential equation which can be integrated easily.

\subsubsection{Reduction with $X_{1}+\mu X_{4}$}

The first possible reduction of the source-free Pulsar equation (\ref{pl.04})
provides the solution to be
\begin{equation}
\Psi_{1}\left(  r,\theta\right)  =r^{\mu}\Sigma\left(  \theta\right)  ,
\label{sc.02}%
\end{equation}
where $\Sigma\left(  \theta\right)  $ satisfies the ordinary differential
equation%
\begin{equation}
\Sigma_{,\theta\theta}+\cot\theta~\Sigma_{,\theta}+\mu\left(  \mu+1\right)
\Sigma=0 \label{sc.03}%
\end{equation}
the closed-form solution of which is given in terms of the Legendre functions as%
\begin{equation}
\Sigma\left(  \theta\right)  =\sigma_{1}P_{\mu}\left(  \cos\theta\right)
+\sigma_{2}Q_{\mu}\left(  \cos\theta\right)  \label{sc.04}%
\end{equation}
in which $P_{\mu}\left(  \theta\right)  ,$ $Q_{\mu}\left(  \theta\right)  $
denote the Legendre functions.

For special values of the parameter $\mu$ the solution (\ref{sc.04}) can be
simplified as follows%
\begin{equation}
\Sigma\left(  \theta\right)  =\sigma_{1}+\sigma_{2}\ln\left(  \frac
{1-\cos\theta}{\sin\theta}\right)  ,~\mu=0, \label{sc.05}%
\end{equation}%
\begin{equation}
\Sigma\left(  \theta\right)  =\sigma_{1}\cos\theta+\sigma_{2}\left(
1-\frac{\cos\theta}{2}\ln\left(  \frac{\cos\theta-1}{\cos\theta+1}\right)
\right)  ,~\mu=1. \label{soll1}%
\end{equation}

It is important to mention that in general the parameter, $\mu$, can be any complex
number and, when it is imaginary, solution (\ref{sc.02}) becomes periodic as
follows $\Psi\left(  r,\theta\right)  =\exp\left(  i|\mu|\ln r\right)
\Sigma\left(  \theta\right)  .~$

Solution (\ref{sc.02}) is well-known in the literature and was derived by
Michel in \cite{pulsar2}. In particular for $\mu=\frac{1}{2}$ solution
(\ref{sc.02}) provides a magnetic field which diverges as the inverse square
root of $r$ such that the total energy of the magnetic field remains finite
at the surface of the star, i.e. when $y=0$. That is a physical condition
which imposes a boundary condition and restricts the free parameters of the solution.

The analytical solutions\ which are presented in the following Sections are
new in the literature, but, as we see, they do not provide explicitly any
law of the form $\Psi\simeq r^{\frac{1}{2}}$.

\subsubsection{Reduction with $X_{2}+\mu X_{4}$}

Consider now reduction with the Lie symmetry vector, $X_{2}+\mu X_{4}$. \ The
invariant solution is calculated in Cartesian coordinates to be
\begin{equation}
\Psi_{2}\left(  y,z\right)  =\Sigma\left(  y\right)  e^{\mu z}, \label{soll2}%
\end{equation}
where the function $\Sigma\left(  y\right)  $ is
\begin{equation}
\Sigma\left(  y\right)  =\sigma_{1}J_{0}\left(  \mu y\right)  +\sigma_{2}%
Y_{0}\left(  \mu y\right)  , \label{soll3}%
\end{equation}
in which $J_{m}\left(  y\right)  ,~Y_{m}\left(  y\right)  $ denote the Bessel
functions of the first and second kind, respectively.

\subsubsection{Reduction with $X_{3}+\mu X_{4}$}

Reduction with the Lie symmetry vector, $X_{3}+\mu X_{4}$, provides
the invariant solution
\begin{equation}
\Psi_{3}\left(  r,\theta\right)  =\frac{1}{r}e^{\mu r\cos\theta}\Sigma\left(
\frac{\sin\theta}{r}\right)  , \label{soll4}%
\end{equation}
where again the function $\Sigma\left(  \frac{\sin\theta}{r}\right)  $ is
expressed n terms of the Bessel functions $J_{m}~$and$~Y_{m}$ as %
\begin{equation}
\Sigma\left(  \frac{\cos\theta}{r}\right)  =\sigma_{1}J_{0}\left(  \mu
\frac{\sin\theta}{r}\right)  +\sigma_{2}Y_{0}\left(  \mu\frac{\sin\theta}%
{r}\right)  .
\end{equation}

\subsubsection{Reduction with $X_{3}+\mu X_{2}$}

The last possible reduction that we can perform in the source-free scenario is
with the use of the Lie symmetry vector, $X_{3}+\mu X_{2}.$ The invariant
solution is calculated to be%
\begin{equation}
\Psi_{4}\left(  r,\theta\right)  =\frac{\Sigma\left(  \sigma\right)  }%
{\sqrt{\mu r^{2}-1}}, \label{soll5}%
\end{equation}
where the new independent variable $\sigma=\sigma\left(  r,\theta\right)  $ is
defined as $\sigma=\frac{r\sin\theta}{\mu r^{2}-1}$. The function
$\Sigma\left(  \sigma\right)  $ satisfies the second-order
differential equation%
\begin{equation}
2\sigma\left(  4\mu\sigma^{2}+1\right)  \Sigma_{,\sigma\sigma}+2\left(
1+12\mu\sigma^{2}\right)  \Sigma_{,\sigma}+6\mu\sigma\Sigma=0,
\end{equation}
the solution of which is expressed in terms of the Legendre functions $P\left(
\sigma\right)  ,~Q\left(  \sigma\right)  ,$ that is,
\begin{equation}
\Sigma\left(  \sigma\right)  =\sigma_{1}P_{-\frac{1}{4}}\left(  8\mu\sigma
^{2}+1\right)  +\sigma_{2}Q_{-\frac{3}{4}}\left(  8\mu\sigma^{2}+1\right)  .
\end{equation}

The source-free equation, (\ref{pl.04}), is linear, a property that follows also
from the existence of the symmetry vectors $X_{4}$ and $X_{\infty}$. Hence
the general solution can be written as a sum of the specific invariant
solutions $\Psi_{1},~\Psi_{2},~\Psi_{3}$ and $\Psi_{4}$ calculated above, over
all the possible values of the free parameters $\mu$ for each solution.
However, the general solution is restricted only when initial/boundary
conditions are applied in the problem.

In the following lines, the reduction process is applied for the remainder of the
cases provided by the Lie symmetry classification.

\subsection{Invariant solutions for linear source}

For the linear source, $F_{C}\left(  \Psi\right)  =F_{1}\Psi$, it is possible
to perform only one reduction with the symmetry vector $X_{2}+\mu X_{4}$.
The invariant solution is calculated in Cartesian coordinates to be%
\begin{equation}
\Psi\left(  y,z\right)  =e^{\mu z}\left(  \Psi_{1}M_{a,0}\left(  2i\mu
y\right)  +\Psi_{2}W_{a,0}\left(  2i\mu y\right)  \right),  \label{soll6}%
\end{equation}
where $\alpha=\frac{iF_{1}}{4\mu};$ and $M_{a,b},$ $W_{a,b}$ are Whittaker functions.~

\subsection{Invariant solutions for power-law source}

For the power-law source, $F_{D}\left(  \Psi\right)  =F_{1}\Psi^{\frac{1}%
{N}+1}$ $\ $, we perform reduction by using the Lie symmetry vector $X_{\left(
N\right)  }$. \ The reduced equation is calculated to be
\begin{equation}
2\sin\theta~\Sigma_{,\theta\theta}+2\cos\theta~\Sigma_{,\theta}+\left(
2N\left(  N-1\right)  \sin\theta-F_{1}\Sigma^{\frac{1}{N}}\right)  \Sigma=0
\label{soll9a}%
\end{equation}
while the solution of the Pulsar equation, (\ref{pl.04}), is expressed as
\begin{equation}
\Psi\left(  r,\theta\right)  =r^{-N}\Sigma\left(  \theta\right)  .
\label{soll7}%
\end{equation}

The reduced equation, (\ref{soll9a}), has been derived before in \cite{pl3,pl4}
and actually the power-law source can describe the magnetic field of the
Pulsar after the surface boundary. More specifically, in \cite{pl4} it was
assumed that, when the source-free axisymmetric pulsar magnetosphere closes,
there exists a boundary condition in order for the solution of the power-law
source to continue to describe the magnetic field. Hence with that assumption it
was found that the value of $N$ is approximately $N\simeq-2.4,$ such that
$\Psi\simeq r^{2.4}$~\cite{pl4}.

It is interesting to comment here that solutions (\ref{sc.02}) and
(\ref{soll7}) were derived before without any knowledge of the symmetries of the
differential equation (\ref{pl.04}). Moreover, those specific invariant
solutions satisfy the boundary conditions imposed by the physics of the
neutron star.

On the other hand, in the coordinates $\left\{  y,z\right\}  $, the reduced
solution can be written equivalently as $\Psi\left(  y,z\right)
=y^{-N}\Lambda\left(  \sigma\right)  $, where $\theta=\arcsin\sigma$, and
$\Lambda\left(  \sigma\right)  $ now satisfies the equation
\begin{equation}
2\sigma\left(  1-\sigma^{2}\right)  \Lambda_{,\sigma\sigma}+2\left(
1+2\sigma^{2}\right)  \Lambda_{,\sigma}+\left(  2\sigma N\left(  N-1\right)
-F_{1}\Lambda^{\frac{1}{N}}\right)  \Lambda=0. \label{eq00}%
\end{equation}
This nonlinear equation does not admit any Lie symmetry and for that we
apply the singularity analysis to study the integrability and write the
analytical solution.

Equation (\ref{eq00}) is a nonautonomous equation.  With the new
change of variables, $\sigma=Y\left(  s\right)  $, $\Lambda\left(
\sigma\right)  =Y_{,s}\left(  s\right)  ,~$we increase the order of the
differential equation, but the new equation is autonomous. We apply the
steps of the ARS algorithm.

We determine the leading-order behaviour to be $Y_{A}\left(  s\right)
=Y_{0}s^{\frac{1}{1+N}}$, for $N\neq-1,$ where $Y_{0}$ is an arbitrary
constant. Hence once expects one of the resonances to be zero.

As far as concerns the resonances they are calculated to be%
\begin{equation}
q_{1}=-1~,~q_{2}=0~,~q_{3}=\frac{2N-1}{1+N},
\end{equation}
which means that the differential equation passes the singularity test and the
\ analytical solutions is expressed by a Right Painlev\'{e} series for
$N\in\left(  -\infty,-1\right)  \cup\left(  \frac{1}{2},+\infty\right)  $ with
step which depends on the value $N$.

In order to perform the consistency test, we select $N=2$ which means that the
third resonance is $q_{3}=1$. Hence the step of the Laurent expansion is
$\frac{1}{3},$ and the~Painlev\'{e} series which describes the solution is
\begin{equation}
Y\left(  s\right)  =Y_{0}s^{\frac{1}{3}}+Y_{1}s^{\frac{2}{3}}+Y_{2}%
s+Y_{3}s^{\frac{4}{3}}+%
{\displaystyle\sum\limits_{I=4}^{\infty}}
Y_{I}s^{\frac{1+I}{3}}.
\end{equation}

The three integration constants are: the position of the singularity and the
coefficients $Y_{0}$ and $Y_{2}$. \ The rest of the coefficients $Y_{J}$ are
functions of $Y_{0}$,~$Y_{2}$. Hence, equation (\ref{soll9a}) is integrable
though the singularity analysis.

However, there exists also a second leading-order behaviour, which is
$Y_{B}\left(  s\right)  =Y_{0}s^{\frac{1}{2-N}},~$with arbitrary $Y_{0}$ and
for all the values of $N$, such that $N\neq2$. The resonances are calculated
to be
\begin{equation}
\bar{q}_{1}=-1~,~\bar{q}_{2}=0~,~\bar{q}_{3}=\frac{2N-1}{N-2}%
\end{equation}
from which we infer that equation (\ref{soll9a}) is integrable.

\subsection{Invariant solutions for the cubic source}

When the power-law source has a cubic law, that is, $F_{E}\left(  \Psi\right)
=F_{1}\Psi^{3}$, then from the symmetry classification we saw that the Pulsar
equation admits an extra Lie symmetry vector field. The reduction with the
vector field $X_{3}$, provides the invariant solution
\begin{equation}
\Psi\left(  r,\theta\right)  =\frac{1}{r}\Sigma\left(  \frac{\sin\theta}%
{r}\right),  \label{soll8}%
\end{equation}
where function $\Sigma$ satisfies the nonlinear differential equation%
\begin{equation}
2\xi\Sigma_{,\xi\xi}+2\Sigma_{,\xi}-F_{1}\Sigma^{3}=0~,~\xi=\frac{\sin\theta
}{r}. \label{th.00}%
\end{equation}

Equation (\ref{th.00}) admits the vector field $X_{\left(  1/2\right)  }$ as Lie
(point) symmetry.  The application of $X_{\left(  1/2\right)  }$ gives%
\begin{equation}
w\left(  \alpha\right)  =\xi^{\frac{3}{2}}\Sigma_{,\xi}~,~\alpha=\xi^{\frac
{1}{2}}\Sigma,
\end{equation}
where now $w\left(  \alpha\right)  $ satisfies the first-order differential
equation%
\begin{equation}
\left(  2+\alpha\right)  w_{,\alpha}-w-F_{0}\alpha^{3}=0,
\end{equation}
which is an Abel's equation of the second kind.

The solution of this Abel's equation cannot be written in a
closed-form.  However, differential equation (\ref{th.00}) can be solved with
the singularity analysis and the generic solution is given in algebraic form.
Hence we apply the ARS algorithm. by firstly making the equation an autonomous
third-order equation with the transformation $\xi=Y\left(  s\right)  $ and
$\Sigma\left(  \xi\right)  =Y\left(  s\right)  _{,s}.$ Equation (\ref{th.00})
is written as%
\begin{equation}
YY_{,s}Y_{,sss}-2Y\left(  Y_{,ss}\right)  ^{2}+\left(  Y_{,s}\right)
^{2}Y_{,ss}-F_{1}\left(  Y_{,s}\right)  ^{6}=0.
\end{equation}

For this equation we perform the new change of coordinates~$Y\left(
s\right)  =\frac{1}{Z\left(  s\right)  },$ where we find that the
leading-order behaviours are $Z\left(  s\right)  =Z_{0}s^{p}$, with $p_{1}=-1$
and $p_{2}=-2.$ In both cases, $Z_{0}$ is an arbitrary constant$.$

For $p_{1}$ the ARS algorithm provides the resonances%
\begin{equation}
q_{1}=-1~,~q_{2}=0~,~q_{3}=\frac{3}{2}%
\end{equation}
which means that the the general solution is given by a Right Painlev\'{e}
expansion with step $\frac{1}{2}$, that is%
\begin{equation}
Z\left(  s\right)  =Z_{0}s^{-1}+Z_{1}s^{-\frac{1}{2}}+Z_{2}+Z_{3}s^{\frac
{1}{2}}+%
{\displaystyle\sum\limits_{I=4}^{\infty}}
Z_{I}s^{-1+\frac{I}{2}} \label{th.01}%
\end{equation}
with free parameters $Z_{0}$~and $Z_{3}$. Note that the third constant of integration
denotes the position of the movable singularity. Finally the
consistency test provides that $Z_{1}=0$, $Z_{2}=-\frac{F_{1}}{2Z_{0}^{2}}%
$,~$Z_{4}=\frac{5\left(  F_{1}\right)  ^{5}}{4Z_{0}^{5}}~etc.$

As far as concerns the second leading-order behaviour, $p_{2}$, we find that the
resonances are
\begin{equation}
\bar{q}_{1}=-1~,~\bar{q}_{2}=-2~,~\bar{q}_{3}=-4
\end{equation}
and, while once expects one of the resonances to be zero, because $Z_{0}$ is
arbitrary, that is not true. Hence the ARS algorithm for the leading-order
term $p_{2}$ fails and solution (\ref{th.01}) is the only solution which can be
constructed by the ARS algorithm.

\begin{figure}[ptb]
\includegraphics[height=7cm]{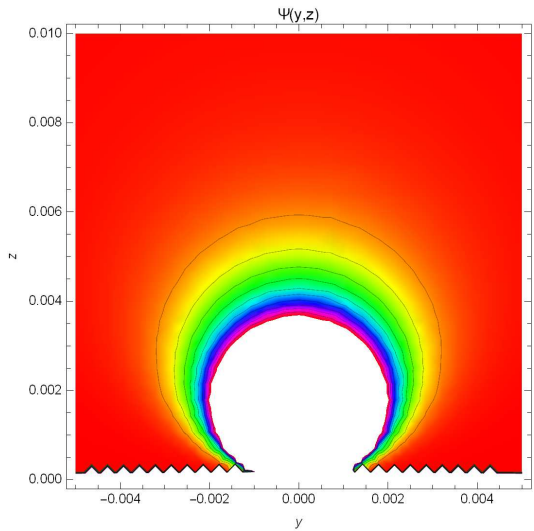}\centering\caption{Density plot of
$\Psi\left(  y,z\right)  $ for the analytical solutions (\ref{soll8}). The
figure is for initial conditions where $\Sigma\left(  0\right)  =0$ and
$\Sigma_{,\xi}\left(  0\right)  >0$.}%
\label{figg01}%
\end{figure}

In Fig. \ref{figg01} the density plot of the invariant solution (\ref{soll8})
is given in space of variables $\left\{  y,z\right\}  $.

For completeness we mention that reduction with the symmetry vector
$X_{\left(  1/2\right)  }$ provides the same solution as that of the
power-law source $F_{D}\left(  \Psi\right)  $ for $N=\frac{1}{2}$.

\subsection{Invariant solutions for exponential source}

Finally for the power-law source, $F_{F}\left(  \Psi\right)  =F_{1}%
e^{-\frac{1}{C}\Psi}$, we apply the invariants of the Lie symmetry vector
field $\bar{X}_{\left(  C\right)  }$, which provide us with the invariant
solution%
\begin{equation}
\Psi\left(  r,\theta\right)  =\ln r^{-C}+\Sigma\left(  \theta\right),
\label{soll9}%
\end{equation}
where $\Sigma\left(  \theta\right)  $, satisfies the nonlinear
second-order differential equation%
\begin{equation}
2\sin\theta~\Sigma_{,\theta\theta}+2\cos\theta~\Sigma_{,\theta}-F_{1}%
e^{-\frac{1}{C}\Sigma}-2C\sin\theta=0.
\end{equation}

As before we prefer to work with the coordinates $\left\{  y,z\right\}  $ \ and
write the invariant solution as%
\begin{equation}
\Psi\left(  y,z\right)  =\ln y^{C}+\Lambda\left(  \sigma\right)  ~,~\text{with
~}\theta=\arcsin\sigma,
\end{equation}
in which function $\Lambda\left(  \sigma\right)  $ satisfies the differential
equation%
\begin{equation}
2\sigma\left(  1-\sigma^{2}\right)  \Lambda_{,\sigma\sigma}+2\left(
1-2\sigma^{2}\right)  \Lambda_{,\sigma}-2C\sigma-F_{1}e^{-\frac{1}{C}\Lambda
}=0. \label{soll10}%
\end{equation}

Equation (\ref{soll10}) has no symmetries and in order to prove the integrability
we apply the ARS algorithm. Indeed, under the change of variables
$\sigma=Y\left(  s\right)  $, $\Lambda\left(  \sigma\right)  =C\ln
Y_{,s}\left(  s\right)  ,$ the leading-order behaviour is calculated to
be~$Y\left(  s\right)  =Y_{0}s^{\frac{1}{1+C}}$, with resonances
\begin{equation}
\bar{q}_{1}=-1,~\bar{q}_{2}=0~,~\bar{q}_{3}=-\frac{1}{1+C}.
\end{equation}

Finally we apply the consistency test of the ARS algorithm for various values
of the parameter $C,$ and we infer that for $C\neq-1$ the differential
equation (\ref{soll10}) passes the singularity test and its solution can be
expressed in terms of a Laurent expansion.

\begin{figure}[ptb]
\includegraphics[height=7cm]{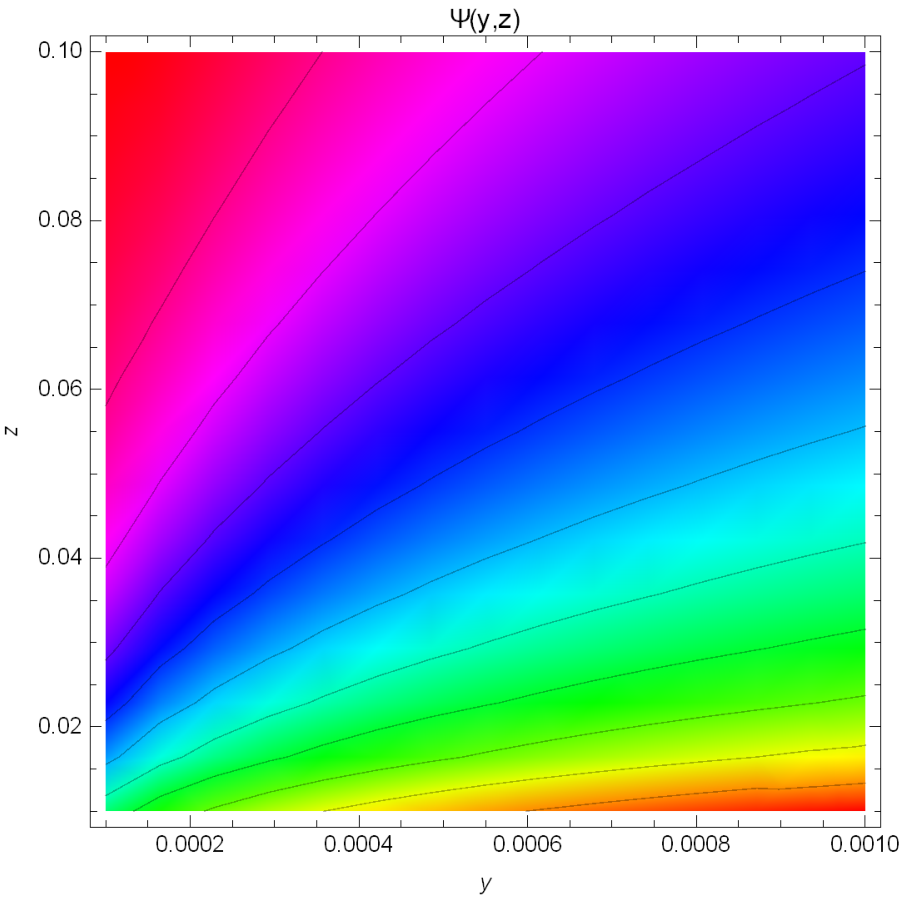}\centering\includegraphics[height=7cm]{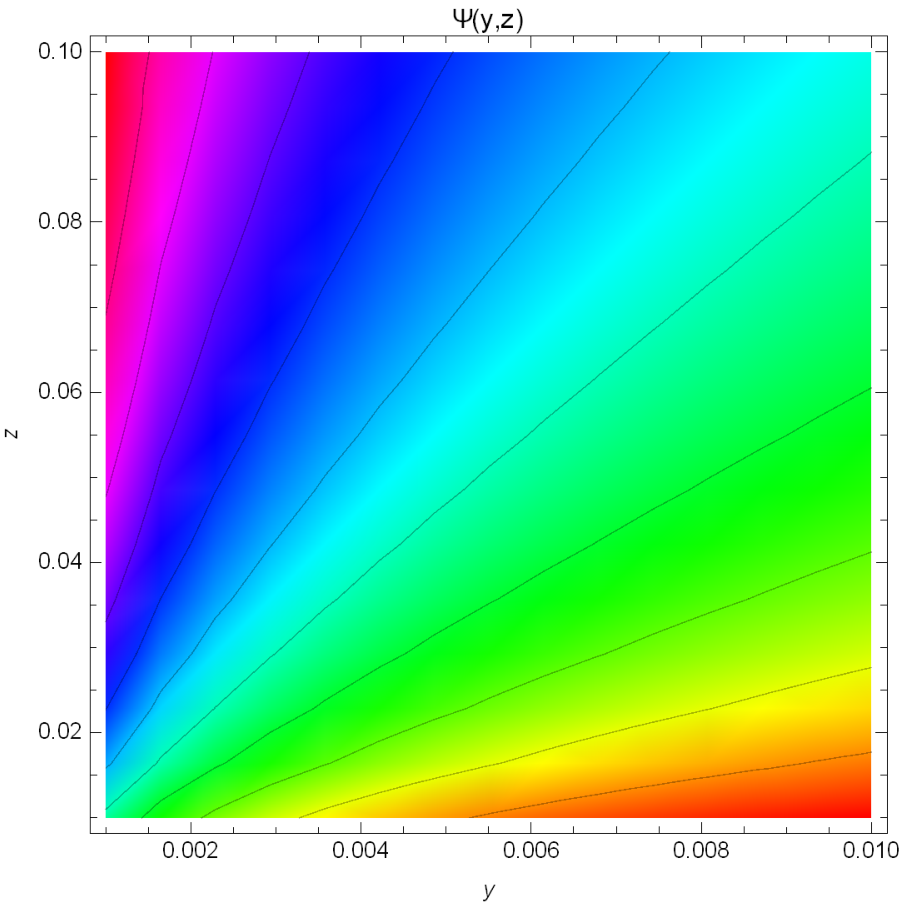}\centering\caption{Density
plot of $\Psi\left(  y,z\right)  $ for the invariant solution (\ref{soll9}).
The plots are for initial conditions $\Sigma\left(  0\right)  =0$ and
$\Sigma_{,\theta}\left(  0\right)  >0$ (left fig.) and $\Sigma_{,\theta
}\left(  0\right)  <0$ (right fig.)}%
\label{figg02}%
\end{figure}

In Fig. \ref{figg02} the density plot of the invariant solution (\ref{soll8})
is given in space of variables $\left\{  y,z\right\}  $.

\section{Conclusions}

\label{sec4a}

In this work we applied two powerful mathematical methods in order to
determine analytical solutions for the Pulsar equation near to the surface of
the neutron star. More specifically we applied the Lie symmetry analysis to
classify the form of the source for the magnetic field in the Pulsar equation
such that the resulting equation admit Lie (point) symmetries, that is, be
invariant under the action of one-parameter point transformations. From the
classification process, we found that the (inner) Pulsar equation can be
invariant under the action of six different Lie algebras.

For each of the vector fields followed by the classification scheme we used
the (zeroth-order) Lie invariants to reduce the number of the independent
variables for the differential equation and write it as an ordinary
differential equation. That equation could be solved in all the cases
with the use of symmetries or with the application of the ARS algorithm. In
particular the ARS algorithm was applied to prove the integrability for some of
the reduced equations and write the analytical solution in a form of Laurent expansion.

The solutions that we derived are asymptotic solutions of the Pulsar equation
(\ref{eq.01}) near to the surface of the neutron star. Only two of the
solutions were derived before in the literature and these Lie invariant solutions
provide a finite magnetic field in the surface of the neutron star. The new
asymptotic solutions can be used as toy-models for the viability of numerical
approximations for the elliptic equation (\ref{eq.01})

In a forthcoming work we wish to study the boundary conditions which should be
satisfied in order that the new Lie invariant solutions be solutions of the
complete problem. Finally the physical implications of those solutions is a
subject for a future study.

\begin{acknowledgments}
AP thanks the University of Athens for the hospitality provided while part of this work
was performed.
\end{acknowledgments}

\end{document}